\documentclass[conference]{IEEEtran}
\IEEEoverridecommandlockouts
\usepackage{cite}
\usepackage{amsmath,amssymb,amsfonts}
\usepackage{algorithmic}
\usepackage{caption}
\usepackage{subcaption}
\usepackage{graphicx}
\usepackage{textcomp}
\usepackage{verbatim}
\usepackage{xcolor}
\usepackage{tikz}
\usepackage{url}
\usepackage{comment}
\usepackage{makecell}
\usetikzlibrary{shapes.geometric, arrows}
\tikzstyle{startstop} = [rectangle, rounded corners,text centered, draw=black]
\tikzstyle{process} = [rectangle, text centered, draw=black]
\tikzstyle{decision} = [diamond, aspect=2, text centered, draw=black]
\tikzstyle{arrow} = [thick,->,>=stealth]

\newcommand{\figurewidth}{0.9\columnwidth}
\newcommand{\figurewidthIII}{0.9\columnwidth}

\def\BibTeX{{\rm B\kern-.05em{\sc i\kern-.025em b}\kern-.08em
    T\kern-.1667em\lower.7ex\hbox{E}\kern-.125emX}}
\begin{document}


\title{ATLAS: \underline{A}I-Native Receiver \underline{T}est-and-Measurement by \underline{L}everaging \underline{A}I-Guided \underline{S}earch}



\author{\IEEEauthorblockN{Mauro Belgiovine$^{\ddagger}$, Suyash Pradhan$^{\star}$, Johannes Lange$^{\dagger}$, Michael Löhning$^{\dagger}$, Kaushik Chowdhury$^{\star}$}
\IEEEauthorblockA{$^{\ddagger}$\textit{NVIDIA Corporation}, Santa Clara, CA, USA.\\$^{\star}$\textit{Chandra Dept. of Electrical and Computer Engineering}, The University of Texas at Austin\\$^{\dagger}$\textit{NI - Test and Measurement Group of Emerson}, Dresden, Germany\\
mbelgiovine@nvidia.com, \{johannes.lange, michael.loehning\}@emerson.com, \{suyash.p, kaushik\}@utexas.edu} \vspace*{-6mm}}

\maketitle

\begin{abstract}


Industry adoption of Artificial Intelligence (AI)-native wireless receivers, or even modular, Machine Learning (ML)-aided wireless signal processing blocks, has been slow. The main concern is the lack of explainability of these trained ML models and the significant risks posed to network functionalities in case of failures, especially since (i) testing on every exhaustive case is infeasible and (ii) the data used for model training may not be available. This paper proposes ATLAS, an AI-guided approach that generates a battery of tests for pre-trained AI-native receiver models and benchmarks the performance against a classical receiver architecture. Using gradient-based optimization, it avoids spanning the exhaustive set of all environment and channel conditions; instead, it generates the next test in 
an online manner to further probe specific configurations that offer the highest risk of failure. We implement and validate our approach by adopting the well-known DeepRx AI-native receiver model as well as a classical receiver using differentiable tensors in NVIDIA's Sionna environment. ATLAS uncovers specific combinations of mobility, channel delay spread, and noise, where fully and partially trained variants of AI-native DeepRx perform suboptimally compared to the classical receivers. 
Our proposed method reduces the number of tests required per failure found by 19\% compared to grid search for a 3-parameters input optimization problem, demonstrating greater efficiency. In contrast, the computational cost of the grid-based approach scales exponentially with the number of variables, making it increasingly impractical for high-dimensional problems.

\end{abstract}


\vspace{-1mm}
\section{Introduction}
\vspace{-1mm}
Next-generation wireless standards are actively exploring incorporating AI/ML approaches for sensing and channel-noise-resilient communication within their wireless transceivers~\cite{9165550,8666641}. Although there have been prototypes and proof-of-concept deployments, there is still a wide gap between fully replacing traditional receivers with a fully AI-native counterpart. The root cause of this gap is the concern that ML models incorporated within an AI-native receiver may not generalize well in various practical environments, leading to catastrophic loss of performance of the network. Thus, without a rigorous and principled methodology for testing and measuring these native AI receivers, operators cannot be assured of their robustness under real-world conditions as e.g. observed on a real-time system-level prototype built with NI's USRP software-defined radio-based research platform \cite{NI_whitepaper}.

\begin{figure}[h]
\centering
\includegraphics[width=1.0\columnwidth]{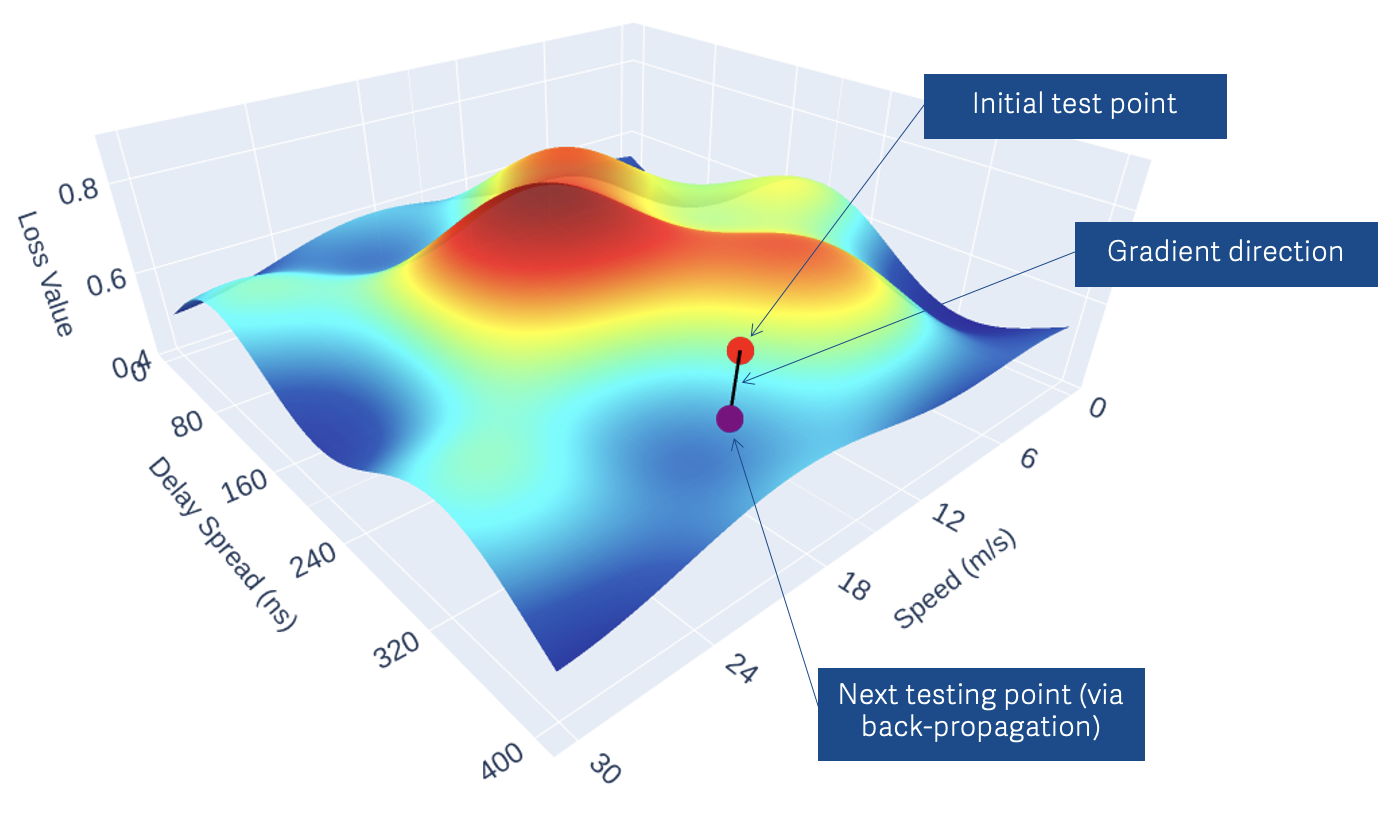}
\caption{Visualization of AI-native gradient-based scenario search for a simplified loss function with two variables (i.e., user terminal's speed and wireless channel delay spread).}
\label{fig:grad_search_concept}
\end{figure}

\begin{figure}[h]
\centering
\includegraphics[width=0.97\columnwidth]{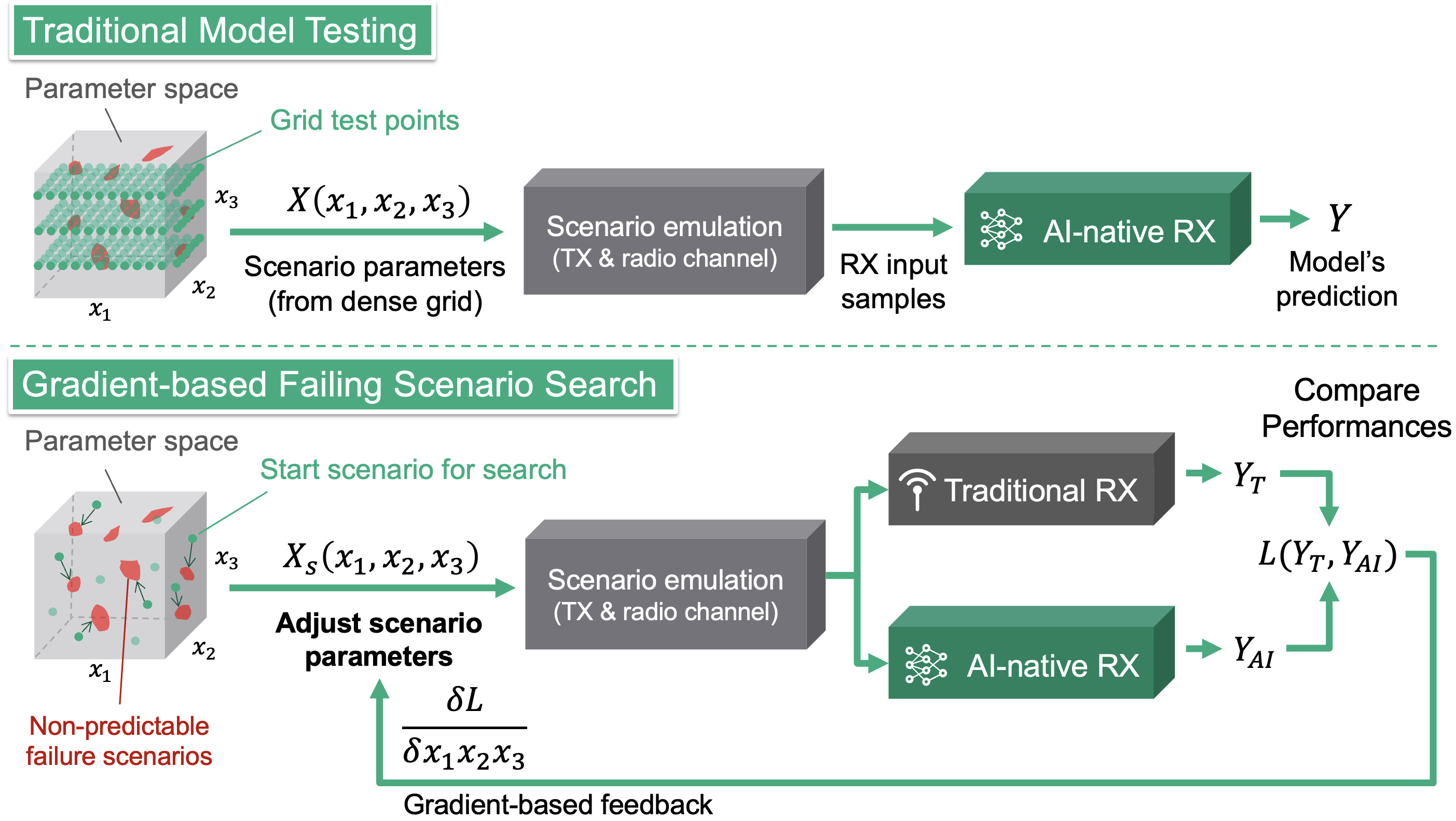}
\caption{Overview of ATLAS gradient-based failure scenario search (bottom) compared to classical grid-based exhaustive search (top). ATLAS employs an iterative optimization loop to fine-tune scenario simulation parameters, crafting input signals that feed into a dual processing pipeline—one path through a differentiable traditional receiver and the other through the AI-native receiver under test. A dedicated loss function then drives parameter updates by maximizing the error of the AI-native receiver and minimizing the traditional one. }
\label{fig:overview}
\end{figure}

\noindent$\bullet$ \textbf{Problem.} 
Traditional receiver designs have well-established performance characteristics based on proven theoretical frameworks. However, there is room for improvement, especially in ensuring low bit error rate (BER) in noisy conditions, where modular ML models have outperformed traditional designs~\cite{9945833, AZARI2022109367}. Replacing individual signal processing blocks with ML counterparts improves the explainability of the AI-native receiver, though the risk remains that ML "black-boxes" may behave unpredictably under certain conditions, especially when not fully exposed to all system variations during training. This risk increases when the entire receiver chain is integrated into a deep neural network, eliminating the modular breakdown for troubleshooting. An example is the DeepRx model for 5G cellular networks~\cite{deeprx}. 
Neither a clear testing methodology exists for Deep learning (DL) based receivers due to the infeasibility of covering all possible wireless scenario conditions, nor have recent advances in eXplainable AI (XAI) \cite{dwivedi2023explainable} yet provided conclusive insights for models designed for the wireless physical layer.

\noindent$\bullet$ \textbf{Solution.} 
ATLAS addresses the challenge of testing AI-native receivers by using a gradient-based optimization approach to efficiently uncover failure cases that traditional grid-based methods or theoretical analyses cannot predict. In Fig.~\ref{fig:grad_search_concept}, we illustrate the procedure of initiating a random test point to validate a DL-based receiver, followed by computing the gradient direction with respect to key scenario variables such as user terminal's speed and wireless channel delay spread. This gradient can then be used to guide the selection of subsequent test points, allowing for a more efficient identification of critical scenarios compared to exhaustive search methods. As shown in Fig. \ref{fig:overview}, conventional testing relies on grid searches over system and environmental configurations (illustrated here along three dimensions), comparing AI-native and traditional receiver performance. Instead, ATLAS leverages the differentiability of DL models and a \textit{differentiable traditional receiver} to compute gradients of key system parameters, guiding an adaptive search via back-propagation to pinpoint failure scenarios. This targeted approach minimizes test data collection and computational overhead while efficiently discovering critical test cases. Implemented using NVIDIA's Sionna \cite{hoydis2022sionna}, a differentiable wireless simulation library built on TensorFlow \cite{abadi2016tensorflow}, ATLAS iteratively refines environmental conditions to generate adversarial inputs that expose weaknesses in AI-native receivers. By integrating feedback from both AI-native and traditional receivers, the framework optimizes signal processing modules and adjusts physical parameters, quickly converging to challenging deployment conditions. This AI-guided test selection not only validates AI-native models but also aids in dataset generation, mitigating the curse of dimensionality with multiple system parameters.

\noindent$\bullet$ \textbf{Contributions.} Our key contributions are as follows:

\begin{itemize}  
    \item 
    We propose a gradient-based approach to efficiently explore the test scenario parameter space for scenario configurations and the derived input signals that degrade the performance of AI-native receiver models.
    \item Our approach leverages the differentiability of DL models and a fully differentiable traditional receiver pipeline, implemented via Sionna, to (i) compare AI-native and traditional receiver performance and (ii) guide parameter exploration to identify unpredictable failure scenarios.  
    \item We evaluate ATLAS using three modified versions of the DeepRx model \cite{deeprx} to assess its effectiveness across different training and parameter configurations.
    
    \item We demonstrate improved efficiency of the proposed approach for an optimization problem based on $3$ channel-dependent variables, achieving a 19\% reduction in the number of tests per number of failures found, compared to grid-based tests. Moreover, we analyze the cost of the two approaches for a varying number of system variables, demonstrating better scalability of proposed solution.
\end{itemize}

\section{Related Work}
\vspace{-1mm}
\noindent$\bullet$ \textbf{Testing Deep Learning models}:
The reliability of DL models has been widely studied, with confidence calibration ensuring alignment between predicted scores and actual accuracy \cite{guo2017calibration, nixon2019measuring}. While effective on image datasets, these methods adjust confidence rather than systematically identifying failure cases, which is crucial for AI-native receivers. XAI techniques like SHAP \cite{SHAP} help interpret unexpected behaviors by highlighting influential input features, aiding structured testing. However, this computes individual contribution of each feature in the input space of a tested model, which scales poorly with the large complex samples of time and frequency-domain wireless signals. Controlled perturbations to input data are frequently used to examine and understand the behavior of complex AI systems. The one-pixel attack \cite{one_pixel} highlights the vulnerability of deep neural networks to low-dimensional adversarial perturbations by using differential evolution to alter just a single pixel in a black-box setting.  Automated white-box testing \cite{test4deep, deepxplore} refines test data samples via back-propagation to expose model weaknesses. Our approach is inspired by these gradient-based methods, which modify the input data of image processing networks to generate additional critical test cases. However, rather than directly perturbing the frequency-domain input data of the neural receiver under test—which could result in unrealistic or non-interpretable samples, we search for critical parameters in the wireless transmission scenario. These parameters correspond to meaningful physical properties of the wireless channel and environment, thereby enabling the generation of realistic and interpretable test cases, enabling efficient testing of AI-native receivers.

\noindent$\bullet$ \textbf{Reliability of AI-native wireless devices}:
Authors in \cite{hoydis2021toward} highlight the challenges of deploying AI-native receivers in real-world wireless environments. \cite{yuan2020transfer} improves adaptability using transfer and meta-learning for beamforming but do not identify failure scenarios. Bayesian approaches, such as those in \cite{bayesian_wireless} and \cite{meta_learning_demod}, model uncertainty to handle outliers and improve demodulator calibration, though they are computationally intensive and rely on accurate priors. 
However, these methods lack active exploration of the test space for adversarial test case generation.

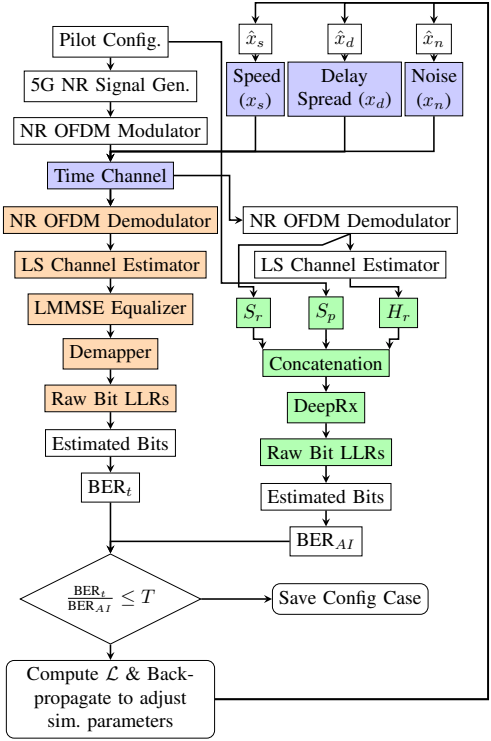
\begin{figure}[t]
\centering
\resizebox {0.75\columnwidth} {!} {
\begin{tikzpicture}[node distance=0.8cm]

\node (pilotconfig) [process] {Pilot Config.};
\node (signal_gen) [process, below of=pilotconfig] {5G NR Signal Gen.};
\node (nrofmdmodulator1) [process, below of=signal_gen] {NR OFDM Modulator};
\node (timechannel) [process, fill=blue!20, below of=nrofmdmodulator1] {Time Channel};
\node (nrofdmdemodulator1) [process, fill=orange!30, below of=timechannel] {NR OFDM Demodulator};
\node (nrofdmdemodulator2) [process, right of=nrofdmdemodulator1, xshift=3.5cm] {NR OFDM Demodulator};
\node (LSchannelestimator2) [process, below of=nrofdmdemodulator2] {LS Channel Estimator};

\node (Y) [process, fill=green!30, below right of=LSchannelestimator2, xshift=-2.3cm, yshift=-0.3cm] {$S_r$};
\node (Xp) [process, fill=green!30, right of=Y, xshift=0.5cm] {$S_p$};
\node (Hr) [process, fill=green!30, right of=Xp, xshift=0.5cm] {$H_r$};
\node (concatenation) [process, below of=Y, xshift=1.3cm, yshift=-0.1cm, fill=green!30] {Concatenation};

\node (deepRx) [process, fill=green!30, below of=concatenation] {DeepRx};
\node (rawbitllrs1) [process, fill=green!30, below of=deepRx] {Raw Bit LLRs};
\node (estimatedbits1) [process, below of=rawbitllrs1] {Estimated Bits};
\node (ber1) [process, below of=estimatedbits1] {$\text{BER}_{AI}$};

\node (LSchannelestimator1) [process, fill=orange!30, below of=nrofdmdemodulator1] {LS Channel Estimator};
\node (lmmseequalizer) [process, fill=orange!30,below of=LSchannelestimator1] {LMMSE Equalizer};
\node (demapper) [process, fill=orange!30, below of=lmmseequalizer] {Demapper};
\node (rawbitllrs2) [process, fill=orange!30, below of=demapper] {Raw Bit LLRs};
\node (estimatedbits2) [process, below of=rawbitllrs2] {Estimated Bits};
\node (ber2) [process, below of=estimatedbits2] {$\text{BER}_{t}$};

\node (decision) [decision, below of=ber2, yshift=-1.2cm] {$\frac{\text{BER}_t}{\text{BER}_{AI}} \leq T$};
\node (saveconfig) [startstop, right of=decision, xshift=3.5cm] {Save Config Case};
\node (adjustsim) [startstop, below of=decision, yshift=-1cm, text width=3.5cm] {Compute $\mathcal{L}$ \& Back-propagate to adjust sim. parameters};

\draw [arrow] (pilotconfig) -- (signal_gen);
\draw [arrow] (signal_gen) -- (nrofmdmodulator1);
\draw [arrow] (nrofmdmodulator1) -- (timechannel);
\draw [arrow] (timechannel) -- (nrofdmdemodulator1);
\draw [arrow] (nrofdmdemodulator2) -- (LSchannelestimator2);
\draw [arrow] (timechannel.east) -- ++(1,0) |- (nrofdmdemodulator2.west);
\draw [arrow] (nrofdmdemodulator1) -- (LSchannelestimator1);


\node (speed) [process, minimum height=0.55cm, text width=0.8cm, fill=blue!20, right of=signal_gen, xshift=1.8cm, yshift=-0.05cm] {Speed ($x_s$)};
\node (delayspread) [process, minimum height=0.55cm, text width=1.8cm, fill=blue!20, right of=speed, xshift=0.8cm] {Delay Spread ($x_d$)};
\node (noise) [process, minimum height=0.55cm, text width=0.8cm, fill=blue!20, right of=delayspread, xshift=0.8cm] {Noise ($x_n$)};
\node (norm_speed) [process, above of=speed,  yshift=0.1cm] {$\hat{x}_s$};
\node (norm_delayspread) [process, above of=delayspread, yshift=0.1cm] {$\hat{x}_d$};
\node (norm_noise) [process, above of=noise, yshift=0.1cm] {$\hat{x}_n$};

\draw [arrow] (speed.south) -- ++(0,-0.63) -| (timechannel.north);
\draw [arrow] (delayspread.south) -- ++(0,-0.63) -| (timechannel.north);
\draw [arrow] (noise.south) -- ++(0,-0.63) -| (timechannel.north);
\draw [arrow] (norm_speed.south) -- (speed.north);
\draw [arrow] (norm_delayspread.south) -- (delayspread.north);
\draw [arrow] (norm_noise.south) -- (noise.north);

\draw [arrow] (nrofdmdemodulator2.south) -- ++(-0.5,-0.2) -| ++(-1.5,0) |- ++(0.1,-0.75) -| (Y.north);
\draw [arrow] (pilotconfig.east) -- ++(0.95,0) |- ++(0,-4.33) -| (Xp.north);
\draw [arrow] (LSchannelestimator2.south) |- ++(0,-0.15) -| (Hr.north);

\draw [arrow] (Y.south) -- ++(0,-0.2) -| (concatenation.north west);
\draw [arrow] (Xp.south) -- ++(0,-0.2) -| (concatenation.north);
\draw [arrow] (Hr.south) -- ++(0,-0.2) -| (concatenation.north east);

\draw [arrow] (concatenation) -- (deepRx);

\draw [arrow] (deepRx) -- (rawbitllrs1);
\draw [arrow] (rawbitllrs1) -- (estimatedbits1);
\draw [arrow] (estimatedbits1) -- (ber1);

\draw [arrow] (timechannel) -- (nrofdmdemodulator1);
\draw [arrow] (LSchannelestimator1) -- (lmmseequalizer);
\draw [arrow] (lmmseequalizer) -- (demapper);
\draw [arrow] (demapper) -- (rawbitllrs2);
\draw [arrow] (rawbitllrs2) -- (estimatedbits2);
\draw [arrow] (estimatedbits2) -- (ber2);

\draw [arrow] (ber2) -- (decision);
\draw [arrow] (ber1.west) -- ++(-3,0) -|  (decision.north);
\draw [arrow] (decision.east) -- ++(0.8,0) |- (saveconfig.west);
\draw [arrow] (decision.south) -- (adjustsim.north);
\draw [arrow] (adjustsim.east) -- ++(4.9,0) -| ++(0,12.5) -- ++(-0.3,0) -| (norm_noise.north); 
\draw [arrow] (adjustsim.east) -- ++(4.9,0) -| ++(0,12.5) -- ++(-0.3,0) -| (norm_delayspread.north); 
\draw [arrow] (adjustsim.east) -- ++(4.9,0) -| ++(0,12.5) -- ++(-0.3,0) -| (norm_speed.north); 

\end{tikzpicture}
}
\caption{
Flow diagram of proposed optimization system, with differentiable traditional 5G transmitter/receiver pipeline (left) and AI-native receiver as the device under test (right). Symbols $\hat{x}_s$, $\hat{x}_d$, $\hat{x}_n$ denote re-parametrized \textit{speed} ($x_s$), \textit{delay spread} ($x_d$), and \textit{noise} ($x_n$) inputs. 
Color-coded blocks represent (1) wireless channel generation (blue), (2) traditional receiver (orange), and (3) DeepRx-based DUT (green). $\mathcal{L}$ is the loss function defined in Eq. \ref{eq:loss}.
}
\label{fig:flowdiagram}
\end{figure}
\vspace{-1mm}
\section{Proposed Gradient-Based Approach}
\vspace{-1mm}
Our proposed method leverages gradient-based optimization to search the 
wireless scenario parameter space for configurations that constitute failure cases in AI-native wireless receivers. Typical applications of gradient-based optimization consist in defining a differentiable loss function used to compute the gradient of a set of variables and using this information to adjust them toward achieving a specific goal. Unlike the traditional application of gradient-based optimization in DL, our approach does not aim to optimize the parameters of the model, but rather the (differentiable) \textit{system parameters} that produce configurations leading to higher output error of the AI-native receivers when compared to traditional architectures.

\vspace{-1mm}

\subsection{Device Under Test (DUT) Description}
\vspace{-1mm}
To illustrate our approach, we consider as the device under test (DUT) an AI-based receiver based on the popular DeepRx~\cite{deeprx} architecture, which consists in a state-of-the-art fully-convolutional neural wireless receiver trained to take in input frequency-domain information, namely (i) received signal $S_r$, (ii) known pilot sequence $S_p$ and (iii) Least-Square coarse channel estimation $H_r$, and produce raw Log-Likelihood Ratio (LLR) estimation of the received bits.The UpLink (UL) signals used for model training and failure search experiments follow 5G NR specifications~\cite{ts38211}, with 14 OFDM symbols (1 ms TTI), 15 kHz sub-carrier spacing, 2-symbol DM-RS pilot configuration, 6 PRBs, and random bit payloads. These frequency-domain signals are modulated to the time domain, transmitted through a wireless channel, and then demodulated back to the frequency domain at the receiver. The DUT performance is measured in terms of Bit-Error-Rate (BER) for a number of different system configurations, which for this work are limited to real-valued wireless channel parameters (i) UE's \textit{speed} $x_s$ (m/s), (ii) \textit{delay spread} $x_d$ (ns) and (iii) channel's \textit{noise power} $x_n$ (dBm), with the latter determining the Signal-to-Noise Ratio (SNR) of simulated transmissions.
Note that, while we focus on channel parameters due to their direct impact on communication performance and desirable differentiable properties, this approach allows to study the impact of any other continuous variables involved in the system. The $\text{BER}_{AI}$ of DeepRx is compared with the $\text{BER}_t$ obtained via a traditional wireless receiver under the same channel conditions, in order to study the differences in their behavior for a given system configuration and receiver input. In the proposed application, we want to identify scenarios where $\text{BER}_{AI}$ exceeds $\text{BER}_t$ by a pre-decided tolerance ratio $T$, defining the following failure scenario trigger condition:
\vspace{-1mm}
\begin{equation}
    \frac{\text{BER}_t}{\text{BER}_{AI}} \leq T
    \label{eq:stop_criteria}
\end{equation}
\subsection{Gradient-based Failing Configuration Search}
\vspace{-1mm}
ATLAS follows an iterative approach, exploring multiple wireless scenarios through search \textit{episodes}. In each episode, we uniformly sample an initial wireless system configuration from all possible configurations expected during deployment. The performance of traditional and AI-native receivers is evaluated using a differentiable dual-receiver pipeline (Fig. \ref{fig:flowdiagram}) to compute the BERs and gradients of the system variables. The system parameters are then adjusted iteratively using the gradients until a failure scenario is identified. Each episode consists of one or more iterations of the following steps:

\begin{enumerate}
    \item \textbf{Wireless Scenario Simulation}:
    For each scenario configuration, we use Sionna simulation blocks to generate a \textit{batch} of 5G NR signals, each passed through an independent simulated wireless channel realization. The respective common scenario parameters are either initially sampled from the scenario parameter space or provided by the gradient-based update during optimization.
    \item \textbf{Parallel Receiver Pipelines}: 
    The resulting batch of receiver input signals is processed by two parallel receiver pipelines: (i) a traditional differentiable pipeline and (ii) a DL-based receiver pipeline. Both pipelines produce a BER value averaged over all the simulated transmissions that serves as a proxy to evaluate the performance gain of AI-based receiver compared to the traditional counterpart. 
    \item \textbf{Failure Case Identification}: Once the BERs are obtained, we evaluate the search stop criteria in Eq. \ref{eq:stop_criteria}. If satisfied, the current scenario configuration is considered to be failing and stored, stopping the current search. Otherwise, the search continues to the next step.
    \item \textbf{Gradient-Based Optimization}: We compute the gradients of studied system-level variables through a loss function that compares the BERs of the two receivers, capturing the impact of system variations on performance differences between  models. 
    We then back-propagate the gradients to adjust the channel configurations, iteratively refining the scenario parameters based on the optimization objective. The process returns to Step 1 and repeats with the updated configuration.  

\end{enumerate}

In order to relate the traditional and AI-based receiver performances and guide the gradient-based exploration in the desired direction, we define a loss function $\mathcal{L}$ as follows:
\vspace{-2mm}
\begin{equation}
    \mathcal{L} = \text{BER}_t - \text{BER}_{AI}
    \label{eq:loss}
\end{equation}
\vspace{-6mm}

Our objective is to \textit{minimize} $\mathcal{L}$ via gradient descent, where $\text{BER}_t$ and $\text{BER}_{AI}$ are \textit{differentiable} functions representing the BER for the traditional and DeepRx-based receivers, respectively, given the same input samples from the differentiable wireless simulator, Sionna. By minimizing $\mathcal{L}$, we force the DeepRx receiver to perform worse than the traditional one ($\text{BER}_{AI} > \text{BER}_t$). For effective performance, we expect $\mathcal{L} < 0$, as $-1 < \text{BER}_t - \text{BER}_{AI} < 1$. This method systematically uncovers weaknesses in ML-based wireless receivers and ensures reliable performance across diverse conditions. Each episode is capped at a maximum of $I$ optimization iterations, with \textit{early-stopping} used to avoid non-productive directions (i.e., when the loss stops decreasing), prompting a new search in a different region. This approach identifies cases where the AI receiver underperforms, highlighting pitfalls and providing configurations to refine future AI solutions.

\vspace{-1mm}
\subsection{Re-parametrization of Optimized System Variables}
\vspace{-1mm}
Optimizing variables with significantly different scales (e.g., speed and delay spread) can cause numerical instability and varying gradient magnitudes, complicating the selection of an appropriate learning rate for smooth updates. To mitigate this issue, we normalize each optimized variable $x$ to a well-conditioned range, typically $[0,1]$, using MinMax normalization:  $\hat{x} = \frac{x - x_{\min}}{x_{\max} - x_{\min}}$ where  $x_{\min}$ and $x_{\max}$ are the known lower and upper bounds of the variable. 
The optimizer computes gradients and updates $\hat{x}$ instead of $x$, ensuring balanced updates. Before using the optimized values, we map them back to the original scale: $x = \hat{x} (x_{\max} - x_{\min}) + x_{\min}$. This ensures stability in a normalized space while maintaining $x_{\text{norm}}$ within $[0,1]$ to prevent out-of-bounds updates.


\begin{table}[t]
    \centering
    \vspace*{0.06in}
    \resizebox{\columnwidth}{!}{
    \begin{tabular}{|l|c|c|c|}
        \hline
        \textbf{Parameter} & \textbf{PTLC} & \textbf{FTLC} & \textbf{FTHC} \\
        \hline
        Modulation & 16 QAM & 64 QAM & QPSK, 16 QAM, 64 QAM\\
        \hline
        Eb/N0 (dB) & $\{0, 1, 2, 3, 18, 19, 20\}$ & $\{0, 2, 4, ..., 20, 22\}$ & $\{0, 2, 4, ..., 20, 22\}$ \\
        \hline
        Delay Spread (ns) & $\{0, 10, 20, 300, 350, 400\}$ & $[0, 400]$ & $[10, 400]$ \\
        \hline
        Speed (m/s) & $\{0, 1, 2, 20, 25, 30\}$ & $[0, 30]$ &  $[0, 30]$ \\
        \hline
        Channel Profile & TDL-D & TDL-D & TDL-B, TDL-C, TDL-D \\
        \hline
        \hline
        \textbf{Model Config.} & \multicolumn{3}{c|}{} \\
        \hline
        \# ResNet Layers & 5 & 5 & 11\\
        \hline
        \# Parameters & 58k & 58k & 700k\\
        \hline
    \end{tabular} 
    }
    \caption{Training parameters for DeepRx-based wireless receiver models studied in this work. Values in curly brackets indicates sets of discrete values, while values in square brackets indicate random sampling in a continuous interval. }
    \label{tab:training_params}
\end{table}

\begin{figure}[h]
     \centering
     \begin{subfigure}[b]{\figurewidthIII}
         \centering
         \includegraphics[width=0.95\columnwidth]{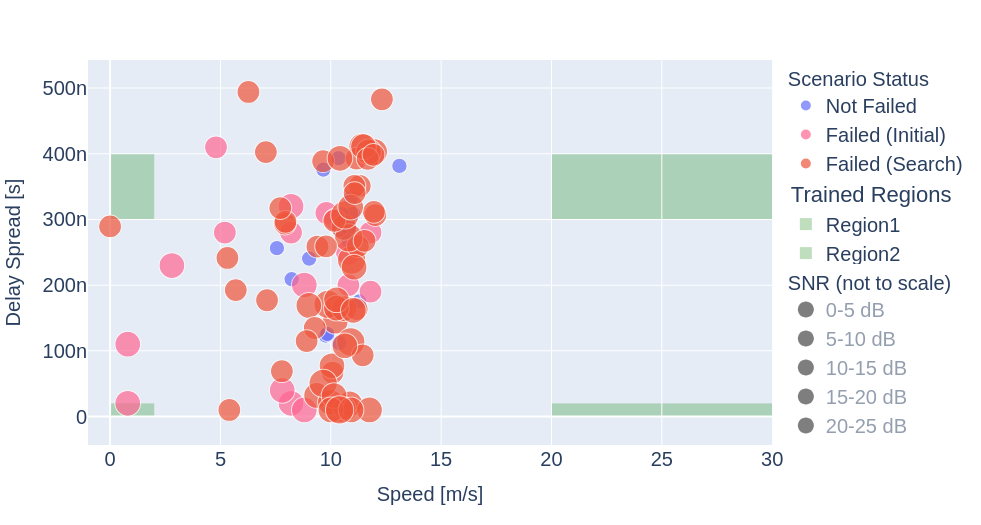}
         \caption{Scenario configurations and test results after gradient-descent failing scenario search across selected channel parameters.}
         \label{fig:ptlc__search_results}
     \end{subfigure}
     \hfill
      \begin{subfigure}[b]{\figurewidthIII}
          \centering
          \includegraphics[width=0.87\columnwidth, trim=0cm 0cm 0cm 0cm, clip]{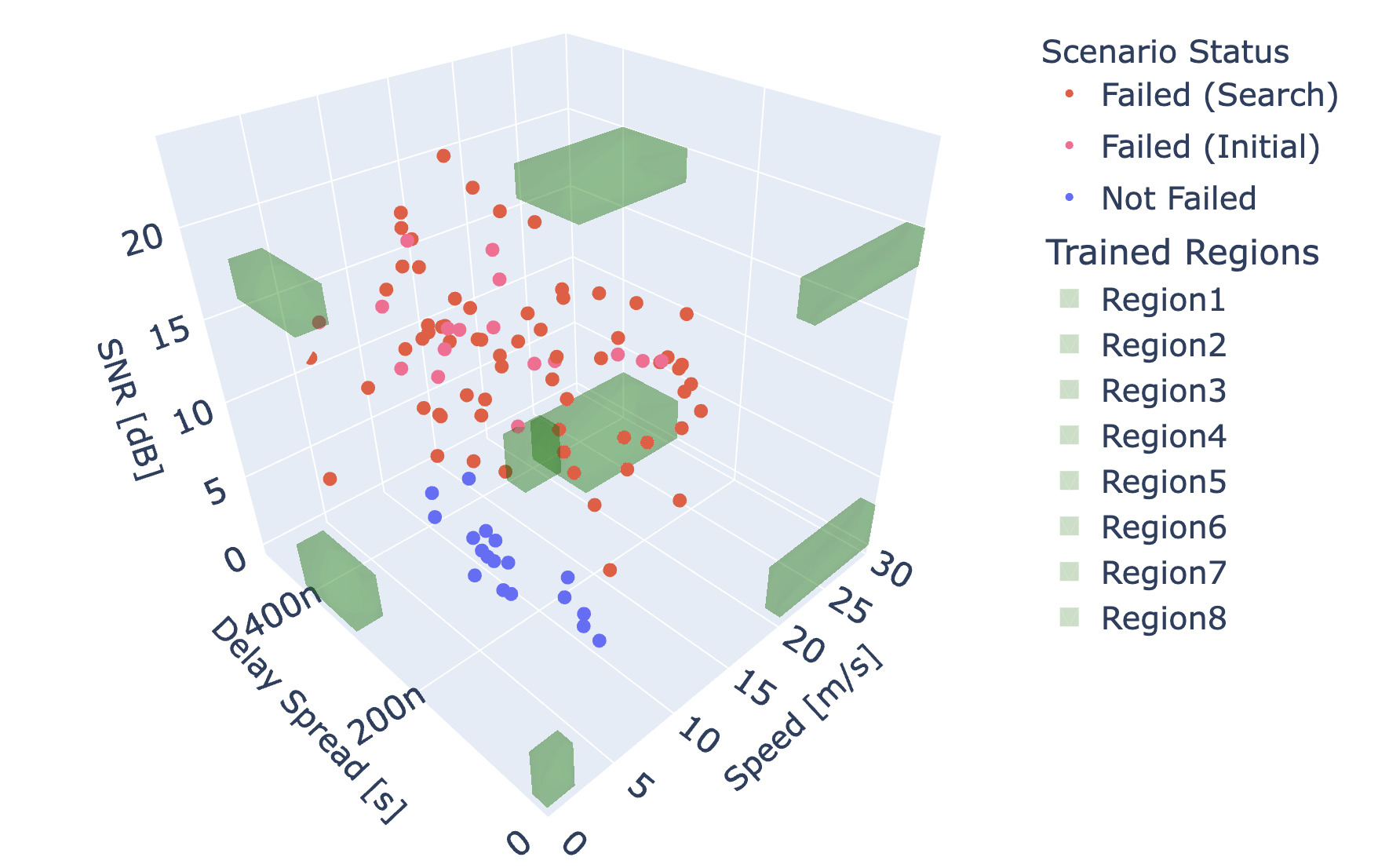}
          \vspace{-1mm}
          \caption{Scenario search results mapped to 3D parameter space}
         
          \label{fig:ptlc__search_results_3d}
      \end{subfigure}
      \hfill
     \begin{subfigure}[b]{\figurewidthIII}
         \centering
         \includegraphics[width=0.95\columnwidth]{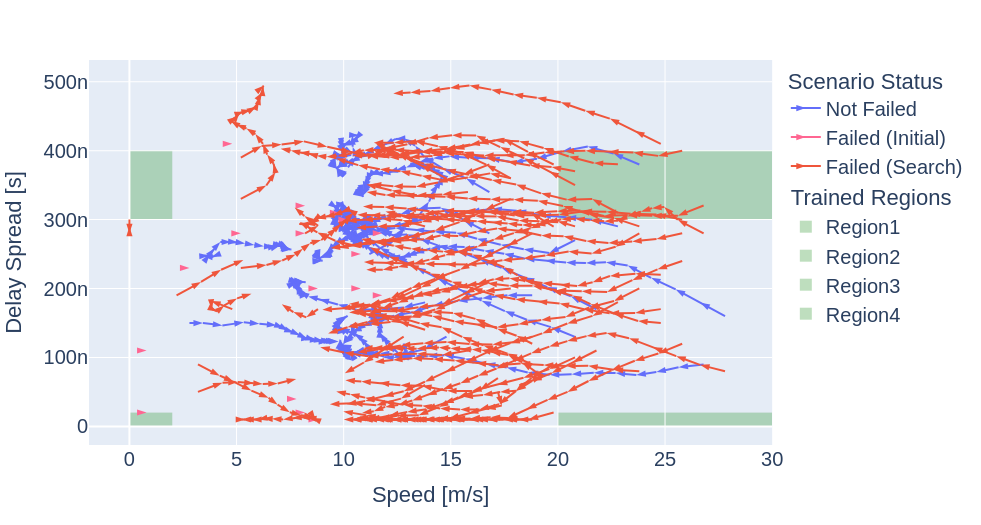}
         \caption{Search trajectories (2D) for each episode across parameter space (arrows indicate the search direction per optimization step)}
         \label{fig:ptlc__search_traj}
     \end{subfigure}
    \caption{Partially Trained, Low Complexity (PTLC) model gradient-based scenario search results. Green areas indicate limited wireless channel parameter value ranges used to train PTLC model sub-optimally.}
     \vspace{-6mm}
    \label{fig:ptlc_search}
\end{figure}

\section{Results from Under-Trained Model}
\vspace{-1mm}
This section demonstrates how ATLAS uncovers weaknesses of AI-native receivers trained under varying conditions. All models were trained using synthetically generated transmissions and bit labels for BER computation with Sionna. We begin with an under-trained version of DeepRx as the DUT, highlighting how ATLAS identifies wireless scenario configurations underrepresented during training. This version, termed \textit{partially trained, low complexity} (PTLC), follows the original model but uses only 5 out of the 11 ResNet layers.

\vspace{-1mm}
\subsection{PTLC Model Training}
\vspace{-1mm}
In this experiment, we under-train the PTLC DeepRx model by generating a dataset with only extreme high and low values for respective wireless channel parameters. The training procedure starts with data generation using Sionna, simulating end-to-end transmission of a 5G NR signal, followed by model training using the generated data. Table \ref{tab:training_params} details the model configuration and parameter ranges for the wireless environment used to generate input samples for model training.
\vspace{-4mm}
\subsection{Failing Scenario Search Results}
\vspace{-1mm}
After training the PTLC model, we perform gradient-based search over \( N_e = 100 \) episodes to identify failure scenarios. Each episode allows up to \( I = 100 \) iterations, starting with an initial learning rate \( l_r = 0.01 \), which is halved if the loss does not improve for 5 epochs. An early stopping criterion is triggered after two such reductions. At each iteration we simulate a batch of 25 scenario realizations (signals and channel instances). 
The start scenario parameters for each episode are sampled from a discrete set evenly distributed across the training ranges, specifically:

\begin{itemize}
    \item Speed (m/s): $\{0, 1, ..., 29, 30\}$;
    \item Delay Spread (ns): $\{10, 20, ..., 380, 400\}$
    \item SNR (dB): $\{5, 10, 15, 20\}$
\end{itemize}
We process signals via traditional and AI pipelines to compute BER for each receiver. If the failure criterion is not met, we use back-propagation with the loss function \( \mathcal{L} \) (see Eq.~\ref{eq:loss}) to compute gradients and update the channel configuration. The batch size is selected based on the available GPU memory (in our case, an NVIDIA A100 with 80GB of VRAM), albeit larger batch sizes generally leading to smoother loss behavior and more stable gradient estimates. Throughout our experiments, we use the highest Modulation and Coding Scheme (MCS) available during training (i.e., 16 QAM) as it is more sensitive to decoding errors. We set a stopping threshold \( T = 0.9 \) (see Eq.~\ref{eq:stop_criteria}), indicating that we terminate the search when the AI-native receiver's BER is at least 10\% higher than that of the traditional receiver. This threshold enables us to identify configurations that are more likely to lead to decoding failures, even though a well-trained AI model is expected to outperform traditional methods across most scenarios. Thus, the choice of \( T \) reflects the desired level of conservativeness in the failure scenario analysis. Fig.~\ref{fig:ptlc_search} summarizes the scenario configurations found via gradient-based search, including 65 failure cases, 17 immediate failures\footnote{Scenarios where the stopping criterion is met before or within a single optimization step.}, and 18 non-failing configurations. 
As shown in Fig.~\ref{fig:ptlc__search_results},  all identified failure cases lie outside the trained channel parameter space as intuitively expected, validating that the search method works. Notably, failures occur only at SNRs $> 10$~dB, indicating that the under-trained AI receiver still outperforms the traditional one in low-SNR regimes. Furthermore, failures cluster around speeds of 10--15~m/s at higher SNRs. These findings confirm that ATLAS not only locates the model's weak points but also highlights regions where such vulnerabilities concentrate. To illustrate how the gradient-based search traverses the parameter space, Fig.~\ref{fig:ptlc__search_traj} shows 2D trajectories over delay spread and speed, demonstrating convergence toward ``blind spot" regions omitted during training.

\vspace{-1mm}
\subsection{Search Results Validation}
\vspace{-1mm}

After identifying failure points via gradient-based exploration, we validate them through an independent run of tests to assess whether these scenarios consistently yield worse BER for the DeepRx model compared to the traditional pipeline. For each identified configuration, we 
average the BER across $1500$ scenario realizations, as memory constraints no longer apply in this offline phase. The same decision criterion in Eq.~\ref{eq:stop_criteria} is used, with results shown for two thresholds: $T = 1.0$ and $T = 0.9$. The higher threshold captures scenarios where DeepRx offers no clear performance gain despite its computational overhead. Such cases are problematic, as DL-based models should justify their cost by consistently outperforming traditional receivers: any configuration in which these failures occur are critical and must be addressed during training.


\begin{table}[t]
\centering
\vspace*{0.06in}
\begin{tabular}{|l|ll|ll|ll|}
\hline
 & \multicolumn{2}{c|}{\textbf{PTLC}} & \multicolumn{2}{c|}{\textbf{FTLC}} & \multicolumn{2}{c|}{\textbf{FTHC}} \\ \hline
\textbf{Grad. Based Search} & \multicolumn{2}{l|}{} & \multicolumn{2}{l|}{} & \multicolumn{2}{l|}{} \\ \hline
Fail (search) & \multicolumn{2}{l|}{65} & \multicolumn{2}{l|}{47} & \multicolumn{2}{l|}{51} \\ \hline
Fail (init) & \multicolumn{2}{l|}{17} & \multicolumn{2}{l|}{4} & \multicolumn{2}{l|}{0} \\ \hline
Not fail & \multicolumn{2}{l|}{18} & \multicolumn{2}{l|}{49} & \multicolumn{2}{l|}{49} \\ \hline
\textbf{Validation} - $T=0.9$ & \multicolumn{1}{l|}{True} & False & \multicolumn{1}{l|}{True} & False & \multicolumn{1}{l|}{True} & False \\ \hline
Fail (search) & \multicolumn{1}{l|}{63} & 2 & \multicolumn{1}{l|}{42} & 5 & \multicolumn{1}{l|}{34} & 17 \\ \hline
Fail (init) & \multicolumn{1}{l|}{17} & 0 & \multicolumn{1}{l|}{4} & 0 & \multicolumn{1}{l|}{0} & 0 \\ \hline
Not Fail & \multicolumn{1}{l|}{18} & 0 & \multicolumn{1}{l|}{49} & 0 & \multicolumn{1}{l|}{49} & 0 \\ \hline
Search Accuracy\textsuperscript{*} & \multicolumn{2}{l|}{0.98} & \multicolumn{2}{l|}{0.95} & \multicolumn{2}{l|}{0.83} \\ \hline
Search Precision\textsuperscript{\dag} & \multicolumn{2}{l|}{0.976} & \multicolumn{2}{l|}{0.902} & \multicolumn{2}{l|}{0.667} \\ \hline
Recall\textsuperscript{\ddag} & \multicolumn{2}{l|}{1.0} & \multicolumn{2}{l|}{1.0} & \multicolumn{2}{l|}{1.0} \\ \hline
\textbf{Validation} - $T=1.0$ & \multicolumn{1}{l|}{True} & False & \multicolumn{1}{l|}{True} & False & \multicolumn{1}{l|}{True} & False \\ \hline
Fail (search) & \multicolumn{1}{l|}{65} & 0 & \multicolumn{1}{l|}{46} & 1 & \multicolumn{1}{l|}{42} & 9 \\ \hline
Fail (init) & \multicolumn{1}{l|}{17} & 0 & \multicolumn{1}{l|}{4} & 0 & \multicolumn{1}{l|}{0} & 0 \\ \hline
Not Fail & \multicolumn{1}{l|}{18} & 0 & \multicolumn{1}{l|}{49} & 0 & \multicolumn{1}{l|}{49} & 0 \\ \hline
Search Accuracy\textsuperscript{*} & \multicolumn{2}{l|}{1.0} & \multicolumn{2}{l|}{0.99} & \multicolumn{2}{l|}{0.91} \\ \hline
Search Precision\textsuperscript{\dag} & \multicolumn{2}{l|}{1.0} & \multicolumn{2}{l|}{0.980} & \multicolumn{2}{l|}{0.823} \\ \hline
Recall\textsuperscript{\ddag} & \multicolumn{2}{l|}{1.0} & \multicolumn{2}{l|}{1.0} & \multicolumn{2}{l|}{1.0} \\ \hline
\end{tabular}
\caption{Performance of proposed gradient-based failure scenario search on PTLC, FTLC and FTHC models with $100$ search episodes applied per model and results validated over $1500$ scenario realizations.
*~=~Ratio of all correct predictions over all predictions. \dag~=~$\frac{\text{TP}}{\text{TP}+\text{FP}}$ (TP = True Positives, FP = False Positives). \ddag~=~$\frac{\text{TP}}{\text{TP}+\text{FN}}$ (FN = False Negatives).}
\label{tab:search_results_all}
\end{table}

\begin{figure}[t]
    \centering

    \begin{subfigure}[b]{0.85\columnwidth}
        \centering
        \includegraphics[width=\linewidth, trim=0cm 0cm 0.5cm 0cm, clip]{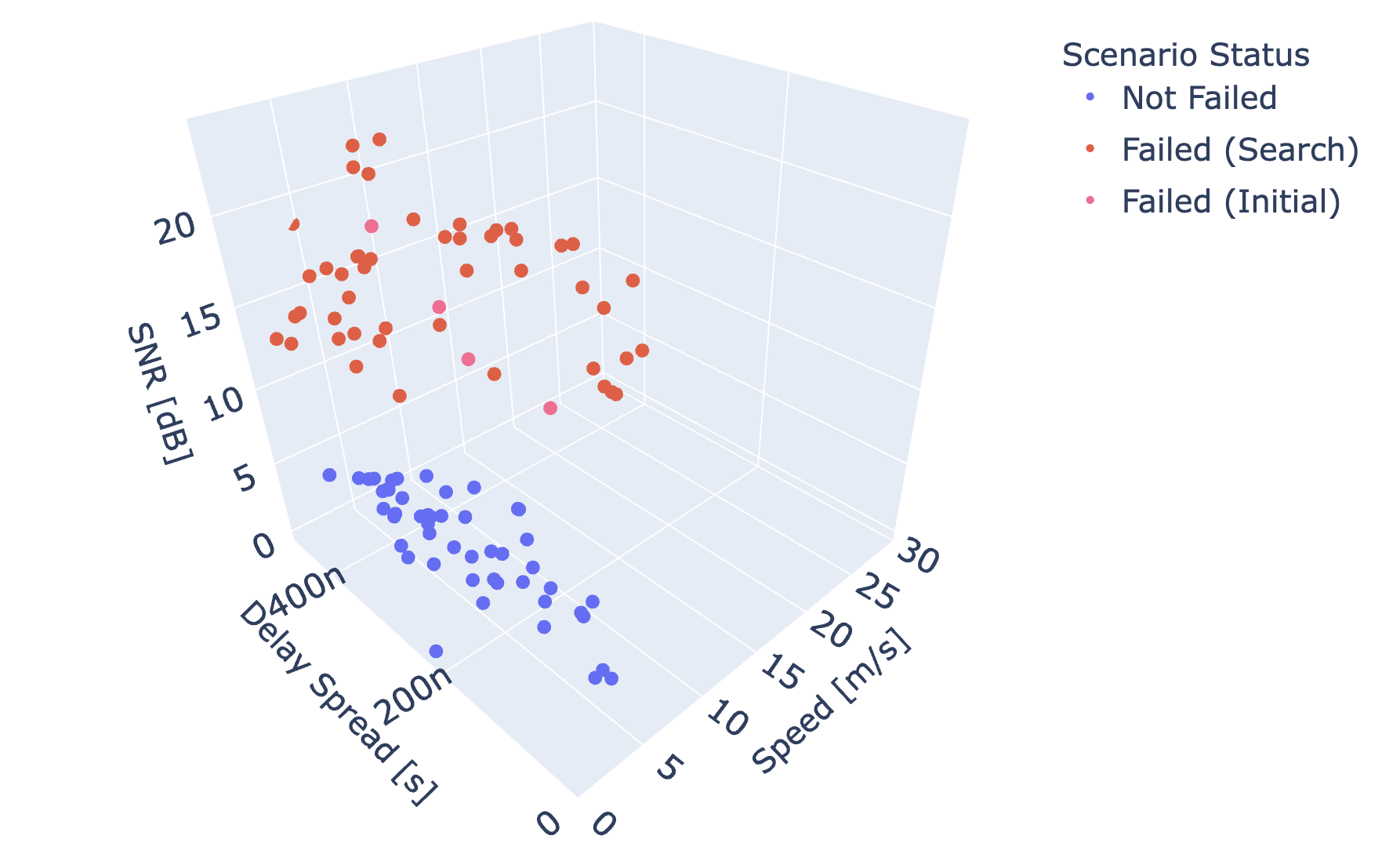}
        \caption{FTLC model}
        \label{fig:ftlc_train_subfig}
    \end{subfigure}

    \vspace{2mm}

    \begin{subfigure}[b]{0.85\columnwidth}
        \centering
        \includegraphics[width=\linewidth, trim=0cm 0cm 0.5cm 0cm, clip]{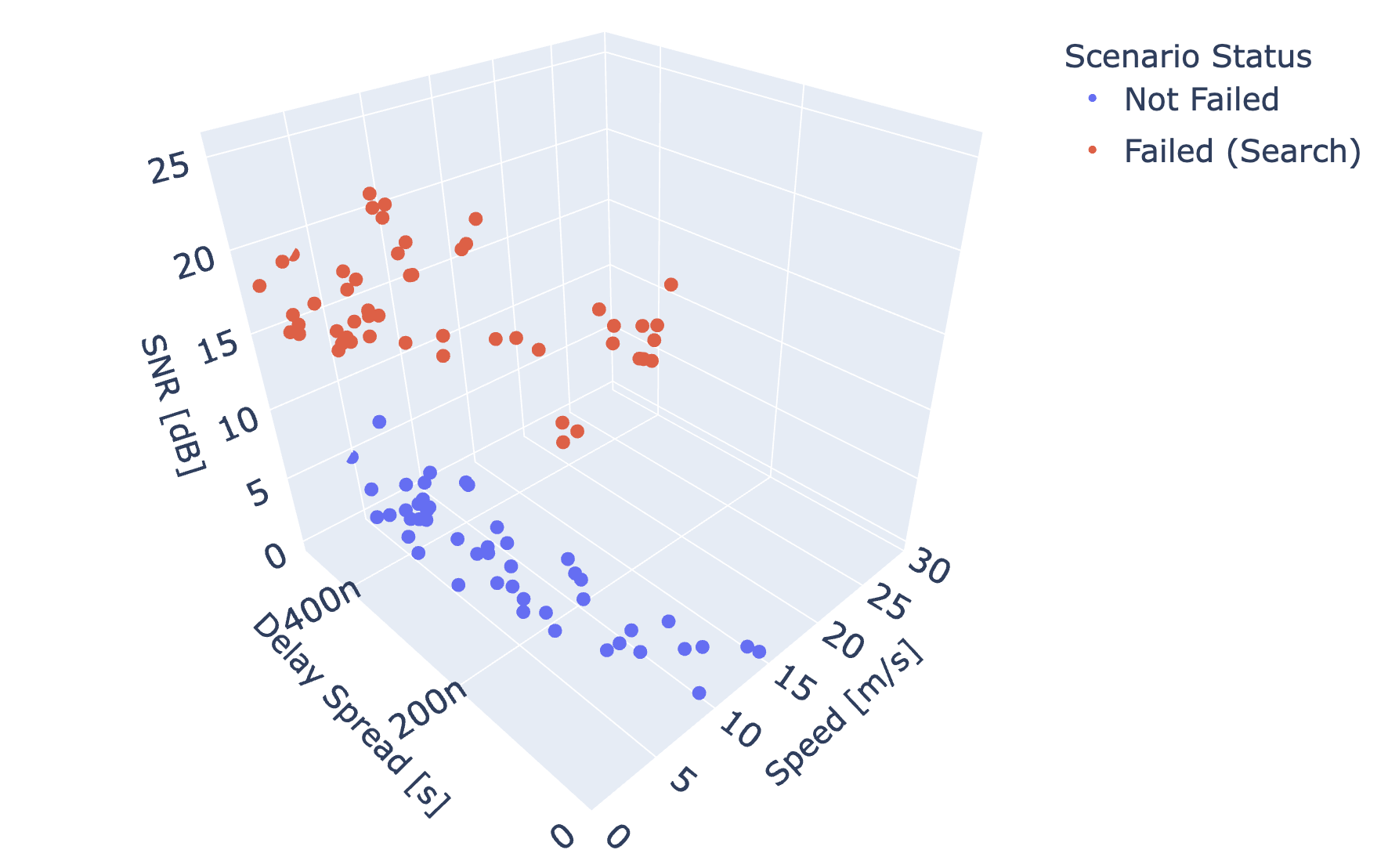}
        \caption{FTHC model}
        \label{fig:fthc_train_subfig}
    \end{subfigure}

    \caption{Search results projected in 3D parameter space for FTLC and FTHC models using gradient-based search.}
    \label{fig:ftlc_fthc_train}
    \vspace{-4mm}
\end{figure}

\vspace{-1mm}
\section{Results from Fully-Trained Models}
\vspace{-1mm}
Once we established the ability of our model to identify configurations of system parameters that affect poorly trained models, we want to test its efficiency on fully trained models with varying complexity, here referenced again in terms of the number of parameters/layers in the convolutional network.
\vspace{-1mm}
\subsection{FTLC and FTHC Models Training}
\vspace{-1mm}
The Fully Trained, Low Complexity (FTLC) and Fully Trained, High Complexity (FTHC) models share the same training procedure but differ in the number of parameters. FTLC has $5$ ResNet blocks (like PTLC), while FTHC uses $11$ ResNet blocks. The aim is to evaluate whether the failure configuration search method can identify failing scenarios in well-trained models and assess the impact of model parameterization (i.e., number of layers and trainable weights). Both models were trained using the procedure outlined for PTLC, with training parameters in Table \ref{tab:training_params}. Similar to PTLC, we test both models at the highest modulation scheme they were trained on (i.e., 64 QAM).

\vspace{-1mm}
\subsection{Failing Scenario Search Results}
\label{sec:fully_trained_search}

Similarly to the PTLC model, we run $100$ search episodes with a maximum of $I = 100$ iterations and early stopping. 
Fig. \ref{fig:ftlc_fthc_train} shows the search results for failing scenarios in FTLC and FTHC models. Both FTLC and FTHC, like PTLC, exhibit a concentration of failing scenarios in higher SNR regions, i.e. SNR $> 10$ dB for FTLC and SNR $> 15$ dB for FTHC. This highlights the denoising capabilities of DL models, which outperform traditional receivers in noisy conditions. However, even fully trained models are affected in higher SNR regions, where the channel effects dominate. This suggests that training datasets for DL-based channel receivers should incorporate more channel samples in higher SNR regions. In terms of failures, both FTLC and FTHC exhibit fewer failure scenarios than PTLC for the same number of episodes (Table \ref{tab:search_results_all}). This is expected, as both models were trained over a broader range of channel parameters, resulting in improved generalization of the model. 
Whereas, FTHC has more trainable weights, allowing it to better adapt to diverse scenarios and reducing rate of failures under unexpected conditions.

\vspace{-1mm}
\subsection{Search Results Validation}
\vspace{-1mm}
While both models exhibit fewer failure points overall, they still show several failures found by the gradient-based search. The results validation shows that the search precision varies with model complexity. For the FTLC model it achieves a precision of $90.2$\% with $T=0.9$ and $98$\% with $T=1.0$. For the FTHC we see lower precision values of $66.7$\% ($T=0.9$) and $82.3$\% ($T=1.0$).
Larger training datasets and increased model parameters improve AI-based receiver robustness but make model verification more challenging. Nevertheless, ATLAS effectively identifies regions in the training space where fully trained models unexpectedly underperform compared to traditional receivers. This helps analyze model weaknesses and refine the bias-variance trade-off for better reliability.

\vspace{-2mm}
\section{Comparison with Grid-Based Testing}
\begin{figure}[t]
\centering
\includegraphics[width=1\columnwidth]{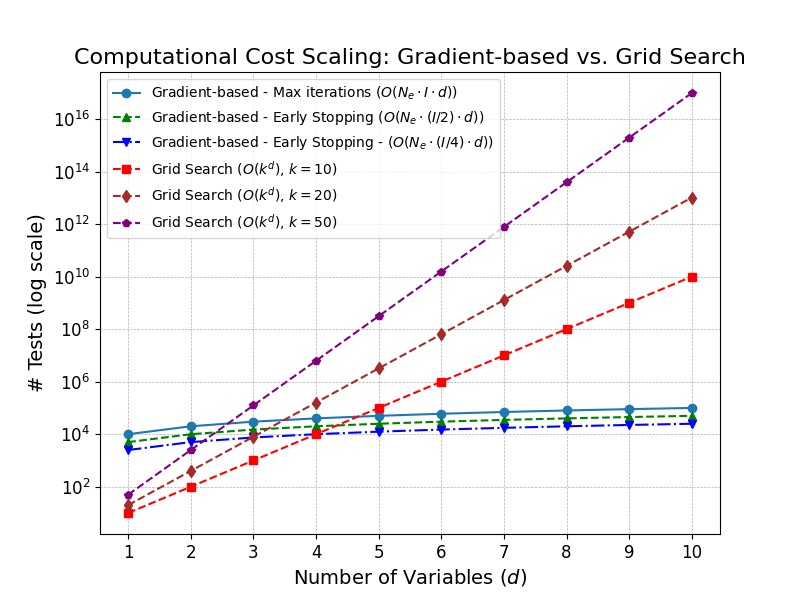}
\caption{Cost scaling comparison of ATLAS vs. grid-based testing as the number of variables $d$ increases. Grid search grows exponentially with $d$, while ATLAS scales linearly.} 
\label{fig:test_cost_comparison}
\end{figure}

\vspace{-1mm}
We compare the efficiency of gradient-based search in ATLAS with conventional grid-based testing using the FTHC model as the DUT. Grid-based testing evaluates uniformly spaced configurations, but its cost grows exponentially with dimension, i.e. $\mathcal{O}(k^d)$ for $k$ points over $d$ parameters, making it impractical in high-dimensional spaces. Although domain knowledge can guide resolution of tests, DL models often react sharply to small environmental changes, making fixed grids unreliable for failure discovery. In contrast, ATLAS samples initial configurations and uses gradients to iteratively navigate toward failure regions without full parameter space coverage. Early-stopping further boosts efficiency by halting unproductive searches. Gradient-based methods like Adam compute only $d$ gradients per step, leading to a linear complexity of $\mathcal{O}(N_e I d)$ across $N_e$ episodes and $I$ iterations each. Table~\ref{tab:grad_vs_grid} compares the cost-efficiency of both approaches, using $\frac{\#\text{Tests}}{\#\text{Failures}}$ as the metric. Here, each optimization step is treated as a single test per the setup in Section~\ref{sec:fully_trained_search}.
For the grid-based approach, we define a uniform sampling space spanning the parameter ranges used during model training, evaluating all points within this discretized configuration space:
\begin{itemize}
    \item Speed (m/s): $[0, 30]$ divided in $25$ intervals;
    \item Delay Spread (ns): $[0, 400]$ divided in $25$ intervals;
    \item SNR (dB): $[0, 22]$ divided in $16$ intervals;
\end{itemize}
for a total of $10$k grid test points. After validating with $T = 1.0$, we observe that ATLAS identifies $42$ failing configurations with an average of $59.8$ tests per failure. In contrast, the grid-based search discovers $135$ failures, but requires an average of $74.1$ tests per failure. This reflects a $19\%$ reduction in the total number of tested configurations using ATLAS, emphasizing the efficiency of a guided search that circumvents exhaustive evaluation of all parameter combinations. 
Fig.~\ref{fig:test_cost_comparison} shows the number of tests required by ATLAS versus grid-based search as the number of parameters $d$ increases. ATLAS scales linearly with $d$ and early-stopping, approximated as an average of $I/2$ and $I/4$ iterations across all search episodes, further reduces cost. In contrast, grid-based methods suffer from exponential growth with $d$ due to the curse of dimensionality. Higher grid granularity ($k$) exacerbates this, making it increasingly difficult for traditional methods to capture DL receivers’ sensitivity to subtle system variations.

\begin{table}[t]
\centering
\vspace*{0.06in}
\begin{tabular}{|l|l|l|}
\hline
 & Gradient-based search & Grid search \\ \hline
\# Episodes & 100 & - \\ \hline
\# Optim. step & 100 (max) & - \\ \hline
\# Tested scenarios & 2511 & 10000 \\ \hline
\# Validated fails ($T=1.0$) & 42 & 135 \\ \hline
\# Tests per failure & 59.8 & 74.1 \\ \hline
\end{tabular}
\caption{Comparison of failing scenario search test efficiency (i.e., ratio of total tests executed over the number of found failing configurations) between proposed gradient-based approach and 
grid-based approach for FTHC model.}
\label{tab:grad_vs_grid}
\vspace{-5mm}
\end{table}
\vspace{-1mm}
\vspace{0mm}
\vspace{-1mm}
\section{Conclusions and Future Work}
\vspace{-1mm}
This work highlights ATLAS’s advantage over traditional grid-based testing by actively targeting failure scenarios instead of exhaustively testing all parameter combinations. Its gradient-based search, enabled by a differentiable traditional and DL-based receiver pipeline, scales efficiently in high-dimensional spaces and uncovers nuanced failure conditions beyond fixed test grids. While simulation results are promising, future work should focus on improving search efficiency and extend applicability to real-world, non-differentiable radio channels via gradient approximation and richer digital twin simulations. Algorithms such as Differential Evolution, as used in~\cite{one_pixel}, or Bayesian Optimization, can offer promising avenues for real system implementation due to their derivative-free approach and better flexibility due to their model-agnostic nature. Building on this, we propose validating ATLAS on a physical testbed, which could be created, e.g., using NI's USRP-based research platform in combination with protocol stacks from Open Air Interface as described in~\cite{NI_whitepaper}.
\vspace{-2mm}
\section{Acknowledgment}
\vspace{-1mm}
This work is partially supported by U.S. National Science Foundation under 2112471.

\vspace{0mm}


\bibliographystyle{unsrt} 
\bibliography{refs}

\end{document}